\documentclass[preprintnumbers,amsmath,amssymbm,prd]{revtex4}
\usepackage{epsfig}
\usepackage{graphicx}
\usepackage{amssymb}

\begin{document}

\title{What can a detected photon with a given gravitational redshift tell us about the maximum density of a compact star?}
\author{Shahar Hod}
\affiliation{The Ruppin Academic Center, Emeq Hefer 40250, Israel}
\affiliation{ } \affiliation{The Hadassah Institute, Jerusalem
91010, Israel}
\date{\today}

\begin{abstract}

\ \ \ Far away observers can in principle bound from below the dimensionless 
maximum-density parameter $\Lambda\equiv4\pi R^2\rho_{\text{max}}$ of a compact star by 
measuring the gravitational redshift factor $z\equiv\nu_{\text{e}}/\nu_{\infty}-1$ of photons that were 
emitted from the {\it surface} of the star: $\Lambda\geq{3\over2}[1-(1+z)^{-2}]$ 
[here $R$ is the radius of the star and $\{\nu_{\text{e}},\nu_{\infty}\}$ 
are respectively the frequency of the emitted light as measured at the location of the emission and by asymptotic observers]. 
However, if photons that were created somewhere {\it inside} the star can make their way out 
and reach the asymptotic observers, then the measured redshift parameter $z$ may not 
determine uniquely the surface properties of the star, thus making the above bound unreliable. 
In the present compact paper we prove that in these cases, in which the creation depth of a detected 
photon is not known to the far away observers, the empirically measured redshift parameter can still 
be used to set a (weaker) lower bound on the dimensionless density parameter of the observed star: 
$\Lambda\geq{3\over2}[1-(1+z)^{-2/3}]$. 
\end{abstract}
\bigskip
\maketitle

\section{Introduction}

The gravitational redshift effect in general relativity \cite{Chan,ShTe} 
implies that electromagnetic waves traveling 
out of a gravitational well (which, for example, may be created by the presence of a compact star) seem to lose energy. 
This physically important phenomenon is reflected in the fact that the frequencies 
of the detected photons as measured by flat-space asymptotic observers are lower than the original frequencies 
of the corresponding photons as determined at the source of the emission (the compact star). 

In spherically symmetric spacetimes the ratio between these two frequencies 
can be expressed in terms of the $tt$-component of the curved line element at the emission site \cite{Chan,ShTe}:
\begin{equation}\label{Eq1}
1+z\equiv{{\nu_{\text{e}}}\over{\nu_{\infty}}}={{1}\over{\sqrt{g_{00}}}}\  ,
\end{equation}
where $\{\nu_{\text{e}},\nu_{\infty}\}$ are respectively the photon frequency as measured at the source of 
emission and the detected frequency as measured by far away asymptotic observers. 
In particular, since the spacetime region just outside the surface of a compact star of mass $M$ and radius $R$ 
is characterized by the simple relation $g_{00}=1-2M/R$ \cite{Noteunits}, the frequencies of 
photons that were created near the surface of the star are gravitationally redshifted according to the simple 
dimensionless ratio \cite{Chan,ShTe} 
\begin{equation}\label{Eq2}
1+z\equiv{{\nu_{\text{e}}}\over{\nu_{\infty}}}=\Big(1-{{2M}\over{R}}\Big)^{-{1\over2}}\  .
\end{equation}

An immediate consequence of the simple relation (\ref{Eq2}) is that the dimensionless 
maximum-density-area parameter
\begin{equation}\label{Eq3}
\Lambda\equiv 4\pi R^2\rho_{\text{max}}\
\end{equation}
of the emitting star can in principle be bounded from below by far away observers who measure the gravitational redshift parameter $z$ of an emitted photon that was created near the {\it surface} of the star: 
\begin{equation}\label{Eq4}
\Lambda\geq{3\over2}[1-(1+z)^{-2}]\  .
\end{equation}
(We have used here the simple inequality $M\leq {4\over3}\pi R^3\rho_{\text{max}}$ for the 
mass of the observed star \cite{NoteK1}).

However, caution should be taken in interpreting observational data of 
stellar emission spectra: one has to take into account the possibility that photons can in principle also 
be created in the non-vacuum region {\it inside} the star, 
where the radial functional behavior of the metric component $g_{00}(r)$ [see Eq. 
(\ref{Eq1})] may not be known to the far away observers. 

In particular, if a photon that was created in some unknown depth inside the {\it non}-vacuum region of 
the star can make its way out, then the empirically measured redshift parameter of the detected photon cannot be used 
by the asymptotic observers to determine uniquely the dimensionless compactness $M/R$ [see Eq. (\ref{Eq2})] 
of the emitting star, thus making the simple lower bound (\ref{Eq4}) unreliable.

The main goal of the present compact paper is to present a compact theorem that reveals the interesting 
fact that in these situations, in which the detected photon was originally 
created in the non-vacuum region inside the star (in some depth which may {\it not} be known 
to the far away asymptotic observers), the observers can still use 
the empirically measured redshift factor of the detected photon in order to derive a lower bound, 
which is albeit weaker than (\ref{Eq4}), on the dimensionless maximum density parameter $\Lambda$ 
of the observed star. 

\section{Description of the system}

We consider a compact star of mass $M$ and radius $R$ whose spherically symmetric 
asymptotically flat spacetime is described, using the Schwarzschild spacetime coordinates, 
by the curved line element \cite{Chan,ShTe}
\begin{equation}\label{Eq5}
ds^2=-e^{-2\delta}\mu dt^2 +\mu^{-1}dr^2+r^2(d\theta^2 +\sin^2\theta d\phi^2)\
\end{equation}
with the radially-dependent metric functions $\mu=\mu(r)$ and $\delta=\delta(r)$. 

The Einstein-matter field equations, $G^{\mu}_{\nu}=8\pi T^{\mu}_{\nu}$, yield the non-linear 
differential relations \cite{May,Hodt1}
\begin{equation}\label{Eq6}
{{d\mu}\over{dr}}=-8\pi r\rho+{{1-\mu}\over{r}}\
\end{equation}
and
\begin{equation}\label{Eq7}
{{d\delta}\over{dr}}=-{{4\pi r(\rho +p)}\over{\mu}}\
\end{equation}
for the metric functions. 
Here 
\begin{equation}\label{Eq8}
\rho\equiv -T^{t}_{t}\ \ \ \ \text{and}\ \ \ \ p\equiv T^{r}_{r}
\end{equation}
are respectively the energy density and radial pressure of the matter fields, 
which are assumed to be non-negative in the interior region of the star and 
respect the dominant energy condition \cite{HawEl,Bond1}:
\begin{equation}\label{Eq9}
0\leq p\leq\rho\ \ \ \ \ \text{for}\ \ \ \ \  r\leq R\  .
\end{equation}
In addition, the energy density and pressure are assumed to vanish outside the 
compact surface of the star:
\begin{equation}\label{Eq10}
\rho=p=0\ \ \ \ \ \text{for}\ \ \ \ \ r>R\  .
\end{equation}

We assume that the spacetime of the compact star is spatially regular, which 
implies the near-origin relations \cite{May,Hodt1}
\begin{equation}\label{Eq11}
\mu(r\to 0)=1+O(r^2)\ \ \ \ {\text{and}}\ \ \ \ \delta(0)<\infty\  .
\end{equation}
In addition, the metric functions of the asymptotically flat spacetime of the star are 
characterized by the functional behaviors \cite{May,Hodt1}
\begin{equation}\label{Eq12}
\mu(r\to\infty) \to 1\ \ \ \ {\text{and}}\ \ \ \ \delta(r\to\infty)\to 0\
\end{equation}
at spatial infinity.

Taking cognizance of the Einstein equation (\ref{Eq6}), one finds 
the simple expression 
\begin{equation}\label{Eq13}
\mu(r)=1-{{2m(r)}\over{r}}\
\end{equation}
for the metric function, where the gravitational mass contained within a sphere of radius $r$ is 
given by the integral relation \cite{May,Hodt1}
\begin{equation}\label{Eq14}
m(r)=4\pi\int_{0}^{r} x^{2} \rho(x)dx\ 
\end{equation}
and is characterized by the boundary condition [see Eq. (\ref{Eq10})]
\begin{equation}\label{Eq15}
m(r=R)=M\  .
\end{equation}

\section{Generic lower bound on the dimensionless maximum density parameter of an optically observed star}

In the present section we shall derive, using {\it analytical} techniques, 
a generic lower bound on the dimensionless maximum-density-area parameter $\Lambda$ of 
optically observed stars. 
In particular, our goal is to derive a model-independent bound which would be valid even in 
situations in which the creation depth inside the star of the asymptotically detected photon is 
not known to the far away observers.

We first point out that Eqs. (\ref{Eq1}), (\ref{Eq5}), and (\ref{Eq13}) yield the functional relation 
\begin{equation}\label{Eq16}
1+z(r_{\text{cr}})={{e^{\delta(r_{\text{cr}})}}
\over{\sqrt{1-{{2m(r_{\text{cr}})}\over{r_{\text{cr}}}}}}}\
\end{equation}
for the asymptotically measured gravitational redshift factor, $z=z(r_{\text{cr}})$, that characterizes a 
photon that originally was created inside the star at a radial distance $r=r_{\text{cr}}$ from its center 
and eventually was detected by the far away asymptotic observers \cite{NoteK2}. 

In order to emphasize the fact that far away observers can in principle determine the redshift factor $z$ 
of an asymptotically detected photon but, in general, they may not have precise knowledge about the exact 
depth (or, equivalently, the exact radius $r=r_{\text{cr}}$) at which the photon was originally created inside the star, we shall henceforth 
use the notation $z$ [instead of $z(r_{\text{cr}})$] for the empirically determined redshift factor 
of an asymptotically detected photon.

Our main goal is to derive a lower bound on the dimensionless density 
parameter $\Lambda$ of the observed star. 
To this end, we shall first prove that the metric function $e^{\delta(r)}/\sqrt{\mu(r)}$, which determines the 
redshift parameter (\ref{Eq16}), is monotonically decreasing inside the star. 
Using the Einstein equations (\ref{Eq6}) and (\ref{Eq7}) with the radial relation (\ref{Eq13}), 
one obtains the gradient relation [see Eq. (\ref{Eq9})]
\begin{equation}\label{Eq17}
{{d\Big[{{e^{\delta(r)}}\over{\sqrt{\mu(r)}}}\Big]}\over{dr}}=-{{e^{\delta(r)}}\over{\sqrt{\mu(r)}}}\cdot
{{{{m(r)}\over{r}}+4\pi r^2p}\over{\mu r}}<0\  ,
\end{equation}
which implies [see Eq. (\ref{Eq11})]
\begin{equation}\label{Eq18}
{\text{max}}_r\Big\{{{e^{\delta(r)}}\over{\sqrt{\mu(r)}}}\Big\}=e^{\delta(0)}\ \ \ \ \ \text{for}\ \ \ \ \ r\in[0,R]\  .
\end{equation}

We shall next derive a useful generic upper bound on the radially-dependent 
dimensionless metric function $\delta(r)$. Using the simple relation [see Eq. (\ref{Eq14})]
\begin{equation}\label{Eq19}
m(r)\leq {{4\pi}\over{3}}r^3\cdot\rho_{\text{max}}\
\end{equation}
for the gravitational mass contained within a sphere of radius $r$, one finds from Eq. (\ref{Eq13}) 
the characteristic inequality
\begin{equation}\label{Eq20}
\mu(r)\geq 1-{{8\pi}\over{3}}r^2\rho_{\text{max}}\  .
\end{equation}

We shall henceforth assume the relation
\begin{equation}\label{Eq21}
\Lambda<{3\over2}
\end{equation}
for the dimensionless density parameter of the star [otherwise we immediately obtain the 
lower bound $\Lambda\geq{3\over2}$, which is actually stronger than the bound (\ref{Eq29}) that we 
shall prove in the present section], which implies the characteristic inequalities [see Eq. (\ref{Eq20})]
\begin{equation}\label{Eq22}
0<\mu(r)\leq 1\  .
\end{equation}

Taking cognizance of Eqs. (\ref{Eq7}), (\ref{Eq9}), and (\ref{Eq20}), one can write 
the following series of inequalities 
\begin{equation}\label{Eq23}
-{{d\delta}\over{dr}}\leq {{8\pi r\rho(r)}\over{\mu(r)}}\leq{{8\pi r\rho_{\text{max}}}\over{1-{{8\pi}
\over{3}}r^2\rho_{\text{max}}}}\
\end{equation}
for the gradient of the metric function in the interior region ($r\leq R$) of the compact star. 
From (\ref{Eq23}) one obtains the integral relation
\begin{equation}\label{Eq24}
-\int^{\delta(R)}_{\delta(r)}d\delta \leq \int^{R}_{r}{{8\pi x\rho_{\text{max}}}\over{1-{{8\pi}
\over{3}}x^2\rho_{\text{max}}}}dx\  .
\end{equation}
Performing the integration in (\ref{Eq24}) with the help of the boundary 
relation [see Eqs. (\ref{Eq7}), (\ref{Eq10}), and (\ref{Eq12})]
\begin{equation}\label{Eq25}
\delta(r=R)=0\  ,
\end{equation}
one finds the characteristic inequality
\begin{equation}\label{Eq26}
\delta(r)\leq {3\over2}\cdot\ln\Big({{3-8\pi r^2\rho_{\text{max}}}\over{3-8\pi R^2\rho_{\text{max}}}}\Big)\
\end{equation}
for the radially-dependent metric function. In particular, one can write the upper bound
\begin{equation}\label{Eq27}
\delta(0)\leq {3\over2}\cdot\ln\Big({{3}\over{3-8\pi R^2\rho_{\text{max}}}}\Big)\
\end{equation}
on the value of the metric function at the center of the star.

Taking cognizance of Eqs. (\ref{Eq3}), (\ref{Eq16}), (\ref{Eq18}), and (\ref{Eq27}), one obtains the dimensionless 
inequalities
\begin{equation}\label{Eq28}
1+z\leq e^{\delta(0)}\leq\Big({{3}\over{3-2\Lambda}}\Big)^{3\over2}\  ,
\end{equation}
which imply the lower bound 
\begin{equation}\label{Eq29}
\Lambda\geq{3\over2}[1-(1+z)^{-2/3}]\
\end{equation}
on the characteristic maximum-density-area parameter of the optically observed compact star. 

We note that the bound (\ref{Eq29}) is weaker than the bound (\ref{Eq4}). 
It is important to emphasize, however, that while the familiar bound (\ref{Eq4}) is only valid if one assumes that the asymptotically detected photon was created near the surface of the star, the analytically derived lower bound (\ref{Eq29}) 
is valid even in situations in which the creation depth of the asymptotically detected photon 
inside the non-vacuum region of the star is not known to us.  

\section{Summary and physical implications}

Determining the value of the composed maximum-density-area 
parameter $\Lambda\equiv4\pi R^2\rho_{\text{max}}$ of highly 
compact stars is a challenging task that may help physicists to determine 
the correct equation of state of highly dense nuclear matter configurations. 
Interestingly, using the bound (\ref{Eq4}), the value of this dimensionless physical parameter can in principle 
be bounded from below by far away observers who measure the gravitational redshift parameter $z$ of an 
asymptotically detected photon that was created near the {\it surface} of the compact star.

In the present paper we have emphasized the fact that, in principle, photons can 
also be created in the non-vacuum region {\it inside} the star, in which case the asymptotically 
measured redshift parameter $z$ may not 
determine uniquely the surface properties of the star, thus making the simple bound (\ref{Eq4}) unreliable. 
 
Motivated by this observation, we have presented a compact theorem that reveals the physically important 
fact that in these situations, in which the detected photon may have been created in the 
non-vacuum region inside the compact star (in some depth which is not necessarily known to the far away 
observers), the asymptotic observers can still use the empirically measured redshift factor of the detected photon 
in order to set an upper bound on the value of the dimensionless density parameter $\Lambda$ that 
characterizes the observed star. 

In particular, using the non-linearly coupled Einstein-matter field equations, 
we have derived the relation [see Eqs. (\ref{Eq3}) and (\ref{Eq29})]
\begin{equation}\label{Eq30}
\Lambda\equiv4\pi R^2\rho_{\text{max}}\geq{3\over2}[1-(1+z)^{-2/3}]\
\end{equation}
between the dimensionless maximum-density-area parameter of an emitting star and the 
gravitational redshift factor that characterizes the detected photon as measured 
by far away asymptotic observers. 

Inspection of the analytically derived lower bound (\ref{Eq30}) reveals the fact that the 
strongest bound on the dimensionless density parameter of the star 
may be deduced by substituting in (\ref{Eq30}) the 
value $z_{\text{max}}$ that characterizes the most redshifted photon which is detected in the emission 
spectrum of the star by the asymptotic observers. 

In order to illustrate the important physical implications of our results, let us assume 
that asymptotic observers detect an emitted photon with say $z=1/2$. 
According to the familiar vacuum relation (\ref{Eq4}), this gravitational redshift factor yields 
the lower bound $\Lambda\geq5/6\simeq0.833$ on the dimensionless density parameter of the emitting star. 
However, the use of the vacuum relation (\ref{Eq4}) may not be justified if the detected photon 
was created in the non-vacuum region inside the star. In these cases one should instead 
rely on the analytically derived density-redshift relation (\ref{Eq30}), 
which implies that the asymptotically detected photon with the property $z=1/2$ 
corresponds to the corrected bound 
$\Lambda\gtrsim0.355$ on the dimensionless maximum-density-area parameter of the observed star.
 
Finally, we would like to emphasize again that the analytically derived lower bound (\ref{Eq30}) on the dimensionless 
density parameter (\ref{Eq3}) of an optically observed star is valid even in situations in which the creation depth 
of the asymptotically detected photon inside the compact star is not known to us.  

\bigskip
\noindent {\bf ACKNOWLEDGMENTS}

This research is supported by the Carmel Science Foundation. I would
like to thank Yael Oren, Arbel M. Ongo, Ayelet B. Lata, and Alona B.
Tea for stimulating discussions.


\begin{thebibliography}{99}
\bibitem{Noteunits} We shall use gravitational units in which $G=c=1$.
\bibitem{NoteK1} Note that the dimensionless density-area parameter $\Lambda=4\pi R^2\rho_{\text{max}}$ 
of the compact star is related to its dimensionless compactness parameter $C=M/R$ by the simple 
inequality $C\leq \Lambda/3$.
\bibitem{NoteK2} It is interesting to note that Eqs. (\ref{Eq3}), (\ref{Eq14}), and (\ref{Eq16}) 
yield the series of inequalities 
$\Lambda/3={4\over3}\pi R^2\rho_{\text{max}}\geq {4\over3}\pi r^2_{\text{cr}}\cdot\rho_{\text{max}}
\geq m(r_{\text{cr}})/r_{\text{cr}}={1\over2}\big\{1-[1+z(r_{\text{cr}})]^{-2}e^{2\delta(r_{\text{cr}})}\big\}$ 
from which one deduces the lower bound 
$\Lambda\geq {3\over2}\big\{1-[1+z(r_{\text{cr}})]^{-2}e^{2\delta(r_{\text{cr}})}\big\}$, where 
$r_{\text{cr}}$ is the radius at which the photon was originally created inside the star. 
As a consistency check we point out that, as should be the case, the relation (\ref{Eq4}) is recovered from this inequality in 
the $r_{\text{cr}}\to R$ limit where $\delta(R)=0$ [see Eqs. (\ref{Eq7}), (\ref{Eq10}), and (\ref{Eq12}) which yield the 
relation $\delta(r=R)=0$ on the surface of the compact star]. 
It is worth emphasizing that far away observers can in principle measure the redshift factor 
of a detected photon but, in general, they may not have precise knowledge about the exact 
radius $r=r_{\text{cr}}$ (the exact depth) at which the photon was originally created inside the star. 
Thus, far away observers may not know the exact value of $\delta(r_{\text{cr}})$ at the creation point 
of the photon inside the star [note that 
Eqs. (\ref{Eq7}), (\ref{Eq9}), (\ref{Eq10}), and (\ref{Eq12}) yield the generic inequality $\delta(r_{\text{cr}})\geq0$]. 
This fact motivates us to derive in the present paper a robust lower bound on the 
density-area parameter $\Lambda$ of the observed star which does not depend on the possibly unknown 
value of the metric function $\delta(r_{\text{cr}})$ at the creation point of the photon 
inside the star.
\end{thebibliography}
\end{document}